\documentclass[a4paper,fleqn,usenatbib]{mnras}

\usepackage[T1]{fontenc}
\usepackage{ae,aecompl}

\usepackage{graphicx}	
\usepackage{amsmath}	
\usepackage{amssymb}	

\usepackage{longtable}
\usepackage{booktabs}
\usepackage{epstopdf}
\usepackage{times}         
\usepackage[flushleft]{threeparttable}

\title[Water Production Rates and Ortho-to-Para Ratios]{\textit{Herschel}/SPIRE Observations of Water Production Rates and Ortho-to-Para Ratios in Comets\thanks{\textit{Herschel} is an ESA observatory with science instruments provided by European-led Principal Investigator consortia and with important participation from NASA.}}

\author[T. G. Wilson et al.]{
Thomas G. Wilson,$^{1}$\thanks{E-mail: tgw@star.ucl.ac.uk}
Jonathan M. C. Rawlings,$^{1}$
and Bruce M. Swinyard$^{1,2}$\thanks{Dedicated to the memory of Prof. Bruce Swinyard.}
\\
$^{1}$Dept. of Physics \& Astronomy, University College London, Gower Street, London WC1E 6BT, UK\\
$^{2}$RAL Space, Science \& Technology Facilities Council, Rutherford Appleton Laboratory, Chilton, Didcot, Oxon OX11 0QX, UK
}

\date{Accepted XXX. Received YYY; in original form ZZZ}

\pubyear{2016}

\begin{document}
\label{firstpage}
\pagerange{\pageref{firstpage}--\pageref{lastpage}}
\maketitle

\begin{abstract}
This paper presents \textit{Herschel}/SPIRE spectroscopic observations of several fundamental rotational ortho- and para-water transitions seen in three Jupiter-family comets and one Oort-cloud comet. Radiative transfer models that include excitation by collisions with neutrals and electrons, and by solar infrared radiation were used to produce synthetic emission line profiles originating in the cometary coma. Ortho-to-para ratios (OPRs) were determined and used to derived water production rates for all comets. Comparisons are made with the water production rates derived using an OPR of 3. 

The OPR of three of the comets in this study are much lower than the statistical equilibrium value of 3, however they agree with observations of comets 1P/Halley and C/2001 A2 (LINEAR), and the protoplanetary disc TW Hydrae. These results provide evidence suggesting that OPR variation is caused by post-sublimation gas-phase nuclear-spin conversion processes. The water production rates of all comets agree with previous work and, in general, decrease with increasing nucleocentric offset. This could be due to a temperature profile, additional water source, or OPR variation in the comae, or model inaccuracies. \\
\end{abstract}




\section{Introduction}
\label{sec:intro}

As comets are formed and spend most of their lifetimes in the outer Solar System they do not undergo significant thermal processing and therefore retain pristine material from the solar protoplanetary disc. Thus studying comets can reveal the history and evolution of the Solar System.
Water is the most abundant volatile in cometary nuclei and its sublimation produces much of the activity seen as comets entering the inner Solar System (heliocentric distance, $r_{\rm{h}},\leq3$\,au).
By studying water in comets comparisons can be made with exoplanetary systems and protoplanetary discs, and a better understanding of planetary formation can be achieved. The ortho-to-para ratio (OPR) of water has been of great interest in recent years in studies looking to understand the history and thermal processing of water in various regions; such as the interstellar medium, star-forming regions, and a protoplanetary disc \citep{Lis2013PhysChem, Choi, Salinas}.
Notably, there is continuing debate as to what fraction of the terrestrial water reservoir was delivered to the Earth by cometary impacts, with key evidence being provided by isotopic ratios such as D/H in cometary ices \citep{Altwegg, Bockelee-Morvan2015, Willacy}. By determining water production rates, $Q_{\rm{H_{2}O}}$, the physical conditions in cometary comae such as temperature, expansion velocity, and excitation conditions can be understood. From constraining the relative abundances of other volatiles compared to water, conditions in the protoplanetary disc and early outer Solar System can be determined. Moreover, the comet to comet $Q_{\rm{H_{2}O}}$ variation (and particularly if there is any distinction between Jupiter-family and Oort-cloud comets) can potentially provide information about the evolution and origin of the cometary ices. Water molecules in cometary comae are excited collisionally by neutrals and electrons, and by solar infrared radiative pumping of fundamental vibration levels. These excitation methods predominantly occur between fundamental rotational levels as cometary comae are typically rotationally cold environments. The majority of the strongest rotational lines, and those mentioned below, are observed in the submillimetre and therefore space-based missions like the \textit{Herschel Space Observatory} provide excellent opportunities for studying the physical properties of comae.

The first direct detection of water in comets was observed in comet 1P/Halley by the Kuiper Airborne Observatory. Fundamental rotational transitions, $2_{\rm{12}}-1_{\rm{01}}$ and $3_{\rm{03}}-2_{\rm{12}}$, were first observed by the Infrared Space Observatory (ISO) in comet C/1995 O1 (Hale--Bopp) \citep{Crovisier1997}. Subsequent space-based missions have detected the fundamental $2_{\rm{12}}-1_{\rm{01}}$ ortho-water line with the \textit{Submillimeter Wave Astronomical Satellite} \citep{Neufeld}, \textit{Odin} \citep{Lecacheux, Biver2007, Biver2009}, and \textit{Herschel} \citep{Hartogh2010, Biver2012}. Observations with \textit{Herschel} have also detected the $2_{\rm{12}}-1_{\rm{01}}$ ortho, and $1_{\rm{11}}-0_{\rm{00}}$ and $2_{\rm{02}}-1_{\rm{11}}$ para-water transitions \citep{de_Val-Borro, Szutowicz, Bockelee-Morvan2012, Lis2013ApJl}. Multiple water rotational lines have been observed by \textit{Rosetta} after the initial detection of the $2_{\rm{12}}-1_{\rm{01}}$ transition \citep{Gulkis}.

For molecules with two protons, each with a nuclear spin angular momentum I = $1/2$, two nuclear spin isomers exist; ortho (I = 1, triplet) and para (I = 0, singlet). Due to the Pauli principle ortho-water exists in rotational states with odd K$_{a} + $K$_{c}$ and para-water with even K$_{a} + $K$_{c}$, as can be seen in Fig.~\ref{fig:rotational}, where K$_{a} + $K$_{c}$ are the projections of the total angular momentum quantum number, J, onto the principal $a$ and $c$ axes. As can be seen in Fig.~\ref{fig:rotational} the lowest ortho- and para-water levels have a rotational energy difference of $34.2$\,K that leads to para-water being more stable in the gas-phase and therefore the OPR can be used as a probe of low temperature regions.

For observations of multiple water rotational lines of both ortho- and para-water transitions the OPR can be determined from equation~(\ref{eq:OPR}), where the OPR is determined as the ratio of sum of all the ortho line intensities divided by their branching ratios, with the sum of all the para line intensities divided by their branching ratios.

\begin{equation}
	\mathrm{OPR} = \dfrac{\sum\limits_{i}I_{\rm{o}}(i)/B_{\rm{o}}(i)}{\sum\limits_{j}I_{\rm{p}}(j)/B_{\rm{p}}(j)}
	\label{eq:OPR}
\end{equation}

where $i$, $j$ indicate individual lines, ortho- and para-intensities are $I_{\rm{o}}(i)$ and $I_{\rm{p}}(j)$, respectively, and $B_{\rm{o}}(i)$ and $B_{\rm{p}}(j)$ are the ortho- and para-branching ratios for each transition.

Typically the OPR in comets is $2.5-3.0$, however multiple comets have been observed with OPR values lower than this and, interestingly, recent studies have reported OPR values in the interstellar medium, star-forming regions, and a protoplanetary disc significantly lower than 3 \citep{Lis2013PhysChem, Choi, Salinas}. From the OPR the nuclear-spin-temperature can be determined \citep*{Mumma1987}.

Nuclear-spin conversion between isomers of isolated molecules occur very rarely due to the weak magnetic interactions between intramolecular nuclear spins. However through hydrogen- or proton-exchange reactions via intermolecular interactions or the mixing of nuclear-spin states via perturbations nuclear-spin conversion can occur. Nuclear-spin conversion can occur collisionally in the gas-phase via the quantum-relaxation model \citep{Hama2013}. If, following a collision, an ortho-water molecule is closer to a para state then coherent mixing of ortho and para states via internal perturbations occurs. A following collision accounts for energy relaxation and the ortho to para conversion is complete.

It has been proposed that in the solid-phase nuclear-spin conversion also occurs through intermolecular spin-magnetic-dipole interactions with neighbouring water molecules on the timescale of $10^{-5}-10^{-4}$\,s. This rapid conversion is due to the rotational energy difference between ortho and para levels decreasing substantially to $5\times10^{-13}$\,K in the solid-phase as there is high barrier for rotation due to hydrogen bonds \citep{Buntkowsky}.

The interpretation of the OPR has been long debated and historically the nuclear-spin-temperature was thought to be indicative of the comet ice formation temperature and therefore comet formation location \citep{Mumma1987}. However, a recent laboratory study has observed that the OPR of both vapor-deposited and in situ-produced water that is sublimated either by thermal desorption at $150$\,K or by photodissociation at 10\,K is equal to a value of 3 \citep{Hama2016}. This study shows that rapid solid-phase nuclear-spin conversion occurs in water ice and normalises the OPR to the statistical equilibrium. Furthermore, the sublimation processes used in the experiment did not alter the OPR. Therefore the OPRs observed in cometary comae are not indicative of the formation temperature of the cometary ices, but instead probe the gas-phase physical conditions in comae.

Although it has been predicted that the collision rate of water with other water molecules, ions, and electrons is too small to induce efficient nuclear-spin conversion in cometary comae \citep{Crovisier1984, Mumma1987}, it has been suggested that comae OPR variation could be due to proton-transfer reactions of water with H$^{+}$ and H$_{3}$O$^{+}$, via water molecule collision with water clusters, or by interactions with ice grains and paramagnetic dust grains in the collisional, fluid, coma regions near the nucleus via the quantum relaxation model described previously \citep{Irvine, Hama2013, MancaTanner}. Therefore nuclear-spin conversion, especially in the low temperature conditions of the coma, needs to be re-examined in order to interpret the observed OPRs.

A recent study of ammonia in 26 comets found a correlation between the ammonia and water nuclear-spin-temperatures potentially suggesting a common process of OPR, and therefore nuclear-spin-temperature, variation.

Observations of 103P/Hartley 2, 10P/Tempel 2, 45P/Honda--Mrkos--Pajdu\v{s}\'{a}kov\'{a}, and C/2009 P1 (Garradd) were undertaken between 2010 July 10 and 2011 October 16 with the Spectral and Photometric Imaging Receiver (SPIRE) \citep{Griffin} instrument onboard \textit{Herschel} \citep{Pilbratt}, in the framework of the \textit{Herschel} Guaranteed Time Key project `Water and related chemistry in the Solar System' (HssO) \citep{Hartogh2009}. The observations are presented in section~\ref{sec:obs} and data analysis including the modeling and results are reported in section~\ref{sec:data}. The main points of the study are discussed in section~\ref{sec:disc} and conclusions are given in section~\ref{sec:conc}.

\section{Observations}
\label{sec:obs}

\begin{figure*}
  \begin{center}
    \includegraphics[width=\linewidth]{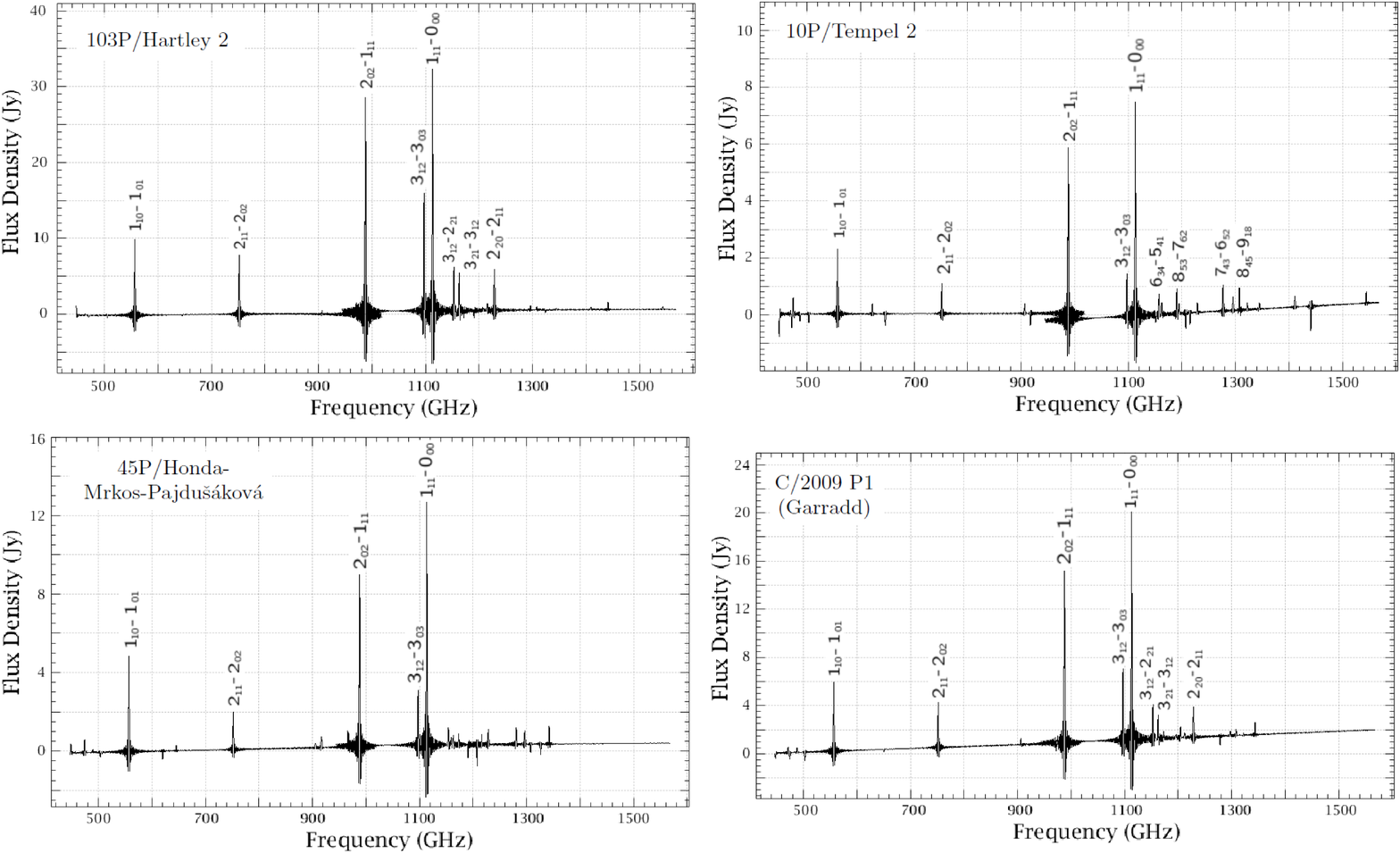}
    \caption{\small{On-nucleus spectra of obtained with SPIRE showing the fundamental rotational water lines and frequencies; $1_{\rm{10}}-1_{\rm{01}}$ ($557$\,GHz, Ortho), $2_{\rm{11}}-2_{\rm{02}}$ ($752$\,GHz, Para), $2_{\rm{02}}-1_{\rm{11}}$ ($988$\,GHz, Para), $3_{\rm{12}}-3_{\rm{03}}$ ($1097$\,GHz, Ortho), $1_{\rm{11}}-0_{\rm{00}}$ ($1113$\,GHz, Para), $3_{\rm{12}}-2_{\rm{21}}$ ($1153$\,GHz, Ortho), $6_{\rm{34}}-5_{\rm{41}}$ ($1158$\,GHz, Ortho), $3_{\rm{21}}-3_{\rm{12}}$ ($1163$\,GHz, Ortho), $8_{\rm{53}}-7_{\rm{62}}$ ($1191$\,GHz, Para), $2_{\rm{20}}-2_{\rm{11}}$ ($1229$\,GHz, Para), $7_{\rm{43}}-6_{\rm{52}}$ ($1278$\,GHz, Ortho), $8_{\rm{45}}-9_{\rm{18}}$ ($1308$\,GHz, Ortho).      
    }}
   \label{fig:spectra} 
  \end{center}
\end{figure*}

The spectra of the comets were acquired with the SPIRE Fourier Transform Spectrometer (FTS) \citep{Swinyard} that covers the spectral range $447-1568$\,GHz with the short (SSW, $191-318\,\mu$m) and long (SLW, $294-671\,\mu$m) wavelength channels. The observations were taken in high resolution mode with a spectral resolution of $\Delta\nu=1.2$\,GHz ($\lambda$/$\Delta\lambda=1000$ at $\lambda=250\,\mu$m). In order to study the comae of the comets the SSW and SLW central bolometers were ignored, effectively creating two rings of SSW beams offset from the nucleus by roughly $33$ and $66$\,arcsec and a ring of SLW beams offset from the nucleus by approximately $51$\,arcsec \citep{Herschel}. The physical offset distance in km is given in Table~\ref{tab:results}. The purpose of studying these SPIRE observations was to determine the OPR and $Q_{\rm{H_{2}O}}$ values in the comae of comets with different formation and evolution conditions at a range of nucleocentric distances.

Data processing was done in HIPE 13.0 using the standard SPIRE scripts. Background subtraction, estimated from the off-axis detectors, and frequency-based mask fitting scripts were also used on Tempel 2, 45P, and C/2009 P1, to produce a better behaved and flatter continuum from which the line intensities were read.   

The JPL HORIZONS system was used to calculate the comet positions and relative motions with respect to \textit{Herschel} and the spectra were corrected for the respective \textit{Herschel}-centric velocities. All observations were taken at offset distances greater than the recombination surface \citep{Bensch}.

The nucleus radius, $r_{\rm{n}}$, and period, $P$, of each comet is given in Table \ref{tab:properties}. The $r_{\rm{h_{obs}}}$, the \textit{Herschel}-comet distance, $\Delta_{\rm{obs}}$, and the time between observation and perihelion, $\Delta{T}_{\rm{obs}}$ where negative values are pre-perihelion and positive values are post-perihelion, when the observations were taken are reported in Table \ref{tab:results}.
 
\subsection{103P/Hartley 2}

103P/Hartley 2 (hereafter Hartley 2) is Jupiter-family comet that passed perihelion on 2010 October 28 at $r_{\rm{h}}=1.059$\,au, a week after making a close approach to the Earth on October 21 at $\Delta=0.12$\,au. Hartley 2 was the target of the EPOXI space mission on 2010 November 4 and was observed by 51 telescopes in a worldwide campaign \citep{Meech}.
One SPIRE FTS observation was taken with a duration of $7002$\,s on 2010 November 9. As can be seen in Fig.~\ref{fig:spectra} several fundamental rotational water emission lines were detected.

\subsection{10P/Tempel 2}

The Jupiter-family comet 10P/Tempel 2 (hereafter Tempel 2) passed perihelion on 2010 July 4 at $r_{\rm{h}}=1.42$\,au shortly before the one SPIRE FTS observation that was taken on 2010 July 10 with a duration of $5650$\,s. Several fundamental rotational water emission lines were detected in Tempel 2 as can be seen in Fig.~\ref{fig:spectra}.

\subsection{45P/Honda--Mrkos--Pajdu\v{s}\'{a}kov\'{a}}

Comet 45P/Honda--Mrkos--Pajdu\v{s}\'{a}kov\'{a} (hereafter 45P) is a Jupiter-family comet that, on 2011 August 15, passed Earth with $\Delta=0.06$\,au before perihelion on 2011 September 28 ($r_{\rm{h}}=0.53$\,au). After the next perihelion passage (2016 December 31) the comet will pass Earth at $\Delta=0.08$\,au on 2017 February 11 providing an opportunity further observation.
Fig.~\ref{fig:spectra} shows the spectra obtained by the SPIRE FTS observation of $4568$\,s on 2011 August 16 with the observed fundamental rotational water lines noted.

\subsection{C/2009 P1 (Garradd)}

Comet C/2009 P1 (Garradd) (hereafter C/2009 P1) is a long period comet originating from the Oort cloud ($i=106^{\circ}$ with respect to the ecliptic). The comet passed perihelion on 2011 December 23 at $r_{\rm{h}}=1.55$\,au and was observed with the SPIRE FTS on 2011 October 16 for $4568$\,s. Fundamental rotational water emission lines can be seen in Fig.~\ref{fig:spectra}.

\section{\textbf{Data Analysis}}
\label{sec:data}

\subsection{Radiative Transfer Model}

\begin{table}
  \centering
  \caption{Orbital and physical properties, and model parameters.} 
  \begin{tabular}{c c c c c} 
    \toprule                   
	& $r_{\rm{n}}$\,(km) & $P$\,(yr) & $\textit{v}_{\rm{exp}}$\,(km\,s$^{\rm{-1}}$) & $\mathit{\beta}_{\rm{H_{\rm{2}}O}}$\,(s$^{\rm{-1}}$) \\
    \midrule\toprule 
     Hartley 2 & $0.7$ & $6.46$ & $0.83$ & $1.08\times10^{-5}$ \\
     Tempel 2 & $5.3$ & $5.36$ & $0.50$ & $1.06\times10^{-5}$ \\                 
     45P & $0.8$ & $5.26$ & $0.75$ & $1.16\times10^{-5}$ \\  
     C/2009 P1 & $<5.6$ & $127,000$ & 0.60 & $1.16\times10^{-5}$\\
    \midrule         
  \end{tabular}
  \label{tab:properties} 
\end{table}

Analysis was carried out using the one-dimensional Accelerated Monte Carlo radiative transfer code; \textsc{rat4com} \citep{Bensch}, adapted from previous work done to generate synthetic water emission spectra \citep{Hogerheijde}. The model includes the excitation of water molecules via collisions with other water molecules and electrons in the inner coma, and by solar infrared pumping of the vibrational bands and fluorescence in the outer coma.

\begin{figure}
  \begin{center}
    \includegraphics[width=\linewidth]{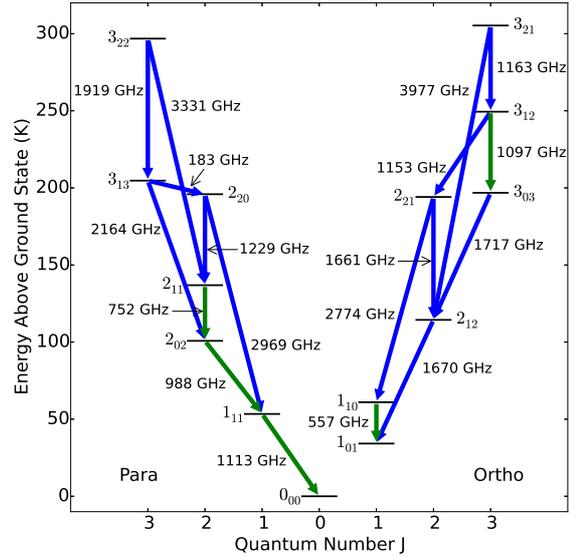}
    \caption{\small{Ortho- and para-water transitions in the \textsc{rat4com} radiative transfer model; unobserved (blue arrows) and observed in all SPIRE detectors (green arrows).}}
    \label{fig:rotational}
  \end{center}
\end{figure}

Previous studies concentrated on ortho-water, considering nine rotational transitions between the seven lowest levels in the ground vibrational state. The updated model also includes the radiative transfer for transitions between the seven lowest levels of para-water. The ortho- and para-water transitions in the updated model can be seen as blue arrows in Fig.~\ref{fig:rotational}, with green arrows representing transitions both in the model and observed in all the SPIRE detectors.

The spherically symmetric Haser distribution \citep{Haser} was used for the radial gas expansion profile and the expansion velocities were assumed to be constant in the cometary comae. This distribution is essentially applicable to a spherically-symmetric, constant expansion velocity ($v_{\rm{exp}}$), outflow (scaling as $1/r^{\rm{2}}$), with an exponential decay term ($\rm{exp}(-r\beta_{\rm{H_2O}}/v_{\rm{exp}}$)) that accounts for photodissociation (rate $\beta_{\rm{H_2O}}$) and ionization by the solar radiation field.

The model parameters that are essentially constrained include $r_{\rm{h}}$, $\Delta$, and scaling factors for $\beta_{\rm{H_{\rm{2}}O}}$ and the ionization rate that take account of $r_{\rm{h}}$ and level of solar activity. Parameters that are reasonably well constrained and for which values from the literature were used include $v_{\rm{exp}}$, the gas kinetic temperature ($T_{\rm{kin}}$), and $\beta_{\rm{H_{\rm{2}}O}}$. The only free parameters are $Q_{\rm{H_{2}O}}$, the electron density scaling factor ($x_{\rm{n_{\rm{e}}}}$), the contact surface scaling factor ($x_{\rm{r_{\rm{e}}}}$), and the number of shells in the radiative transfer calculations.

At the nucleocentric distances observed a constant $T_{\rm{kin}}=40$\,K is assumed as previous work has shown this value to be a good approximation \citep{Combi1999}. Observations of the $1_{\rm{10}}-1_{\rm{01}}$ ($557$\,GHz) water line in other comets have constrained $x_{\rm{n}_{\rm{e}}}$ to $0.2$ \citep{Biver1997, Biver2007, Hartogh2010}. Previous observations of 1P/Halley have shown that $x_{\rm{r_{\rm{e}}}}$, a scaling factor that determines the boundary radius between collisional excitation predominantly caused by water-water and water-electron, is equal to unity \citep{Balsiger, Festou}. The number of shells was set at $1500$ based on recommendations \citep{Bensch}. Adopted values for the other model parameters mentioned above are given in Table~\ref{tab:properties}.

Although this model makes some simplistic assumptions such as spherical symmetry of the outgassing, and in the irradiation and photochemistry of the water, the omission of vibrationally excited levels, and radiative transfer at infrared wavelengths a number of comet observations have been modeled with considerable success \citep{Zakharov, Hartogh2010, Bockelee-Morvan2012}.  

\subsection{Ortho-to-Para Ratios}

As mentioned in section~\ref{sec:intro} due to the multiple transitions seen in the observations OPR values can be calculated using equation~(\ref{eq:OPR}) and knowledge of the OPR of a comet can lead to the determination of the nuclear-spin-temperature.  Note that many of the upper levels of these transitions are populated from higher levels that are not included in the radiative transfer model. The excitation mechanism(s) for these levels is not described in the models. The observed transitions are labeled as ortho- or para-water in Fig.~\ref{fig:spectra} and the calculated OPR values can be seen in Table~\ref{tab:results}. For each comet the OPR values determined agreed over the multiple offsets mentioned in section~\ref{sec:obs} and therefore the values presented are an average over the observed nucleocentric distances. It should be noted that at these offsets correspond to nucleocentric distances of $\sim1000-10000$\,km and the very inner comae of all comets are not observed.

As can be seen in Table~\ref{tab:comparison}, Hartley 2 has been studied previously and a wide range of OPR values have been determined \citep{Crovisier1999, DelloRusso2011, Mumma2011, Bonev2013, Kawakita}. The calculated value presented below agrees with the literature and gives a nuclear-spin-temperature of $\approx28$\,K.

By comparing the OPR values for Tempel 2 in Table~\ref{tab:comparison}, the OPR value reported in this study is a factor of two lower than previously seen \citep{Paganini2012Ic} and results in a nuclear-spin-temperature of $\approx20$\,K. As both sets of observations were taken at similar $r_{\rm{h}}$ this suggests an OPR variation in the coma over the $16$\,d between observations. While this is an extreme case of OPR variation a previous example has been seen. The OPR of comet C/2001 A2 (LINEAR) (hereafter C/2001 A2) was observed to decrease from $2.5$ to $1.8$ in $24$\,h \citep{DelloRusso2005}. Possible causes of OPR variation are presented in section~\ref{sec:intro} and discussed in section~\ref{sec:opr_var}.

For both 45P and C/2009 P1 no previous OPR determinations have been calculated. The determined OPR values for these comets reported in this study correspond to a nuclear-spin-temperature of $\approx22$\,K and $\approx18$\,K respectively. While the OPR values presented in Table~\ref{tab:results} are lower than previously observed in comets, they agree with previous OPR values seen in 1P/Halley and C/2001 A2 \citep{Mumma1988, DelloRusso2005}. 

\subsection{Water Production Rates}

\begin{table*}
  \centering
  \caption{Coma averaged OPR values for the four \textit{Herschel}/SPIRE comets. $Q_{\rm{H_{2}O}}$ determined for various lines and nucleocentric offsets. Using; (i) the calculated OPR values and (ii) an OPR of 3.} 
  \begin{tabular}{c c c c c c c c c c c c} 
    \toprule                   
	Comet & Start Date (UT) & $\Delta{T}_{\rm{obs}}$ & $r_{\rm{h_{obs}}}$ & $\Delta_{\rm{obs}}$ & OPR & Transition & $\nu_{\rm{ij}}$ & Ortho &Offset & \multicolumn{2}{c}{$Q_{\rm{H_{2}O}}$ ($10^{27}$\,s$^{-1}$)} \\
     & yyyy/mm/dd.dd & (d) & (au) & (au) &  &  & (GHz) & / Para & (km) & (i) & (ii) \\
    \bottomrule\toprule                  
    Hartley 2 & 2010/11/09.03 & $11.77$ & $1.071$ & $0.176$ & $2.44\pm0.71$ & $1_{10}-1_{01}$ & $557$ & Ortho & $6400$ & $3.83\pm0.12$ & $3.61\pm0.17$ \\
      \cmidrule{7-12}
      &  &  &  &  &  & $2_{11}-2_{02}$ & $752$ & Para & $6400$ & $1.89\pm0.11$ & $2.19\pm0.15$ \\
      \cmidrule{7-12}
      &  &  &  &  &  & $2_{02}-1_{11}$ & $988$ & Para & $4100$ & $6.43\pm0.13$  & $7.49\pm0.21$ \\
      &  &  &  &  &  &  &  &  & $6400$ & $5.68\pm0.11$ & $6.57\pm0.12$ \\
      &  &  &  &  &  &  &  &  & $8200$ & $4.09\pm0.15$ & $4.75\pm0.20$ \\
      \cmidrule{7-12}
      &  &  &  &  &  & $3_{12}-3_{03}$ & $1097$ & Ortho & $4100$ & $0.55\pm0.06$ & $0.55\pm0.06$ \\
      &  &  &  &  &  &  &  &  & $8200$ & $0.35\pm0.04$ & $0.35\pm0.04$ \\
      \cmidrule{7-12}
      &  &  &  &  &  & $1_{11}-0_{00}$ & $1113$ & Para & $4100$ & $7.46\pm0.13$ & $8.67\pm0.19$ \\
      &  &  &  &  &  &  &  &  & $8200$ & $5.57\pm0.14$ & $6.44\pm0.20$ \\      
    \midrule
    Tempel 2 & 2010/07/10.93 & $6.03$ & $1.424$ & $0.732$ & $1.59\pm0.23$ & $1_{10}-1_{01}$ & $557$ & Ortho & $27000$ & $11.6\pm1.7$ & $9.5\pm1.4$ \\
      \cmidrule{7-12}
      &  &  &  &  &  & $2_{11}-2_{02}$ & $752$ & Para & $27000$ & $3.8\pm1.2$ & $5.7\pm1.8$ \\
      \cmidrule{7-12}
      &  &  &  &  &  & $2_{02}-1_{11}$ & $988$ & Para & $17000$ & $9.5\pm1.2$ & $14.5\pm1.8$ \\
      &  &  &  &  &  &  &  &  & $27000$ & $5.2\pm0.6$ & $8.2\pm0.9$ \\
      &  &  &  &  &  &  &  &  & $35000$ & $10.2\pm1.7$ & $15.8\pm2.5$ \\
      \cmidrule{7-12}
      &  &  &  &  &  & $3_{12}-3_{03}$ & $1097$ & Ortho & $17000$ & $2.9\pm1.3$ & $2.3\pm1.1$ \\
      &  &  &  &  &  &  &  &  & $35000$ & $2.2\pm1.0$ & $1.8\pm0.8$ \\
      \cmidrule{7-12}
      &  &  &  &  &  & $1_{11}-0_{00}$ & $1113$ & Para & $17000$ & $8.6\pm0.9$ & $13.1\pm1.3$ \\
      &  &  &  &  &  &  &  &  & $35000$ & $6.0\pm1.0$ & $9.2\pm1.5$ \\   
    \midrule
    45P & 2011/08/16.13 & $-44.65$ & $1.002$ & $0.061$ & $2.00\pm0.30$ & $1_{10}-1_{01}$ & $557$ & Ortho & $2400$ & $1.22\pm0.11$ & $1.08\pm0.09$ \\
      \cmidrule{7-12}
      &  &  &  &  &  & $2_{11}-2_{02}$ & $752$ & Para & $2400$ & $0.37\pm0.09$ & $0.49\pm0.12$ \\
      \cmidrule{7-12}
      &  &  &  &  &  & $2_{02}-1_{11}$ & $988$ & Para & $1500$ & $1.25\pm0.10$ & $1.67\pm0.14$ \\
      &  &  &  &  &  &  &  &  & $2400$ & $0.94\pm0.05$ & $1.26\pm0.06$ \\
      &  &  &  &  &  &  &  &  & $3100$ & $0.68\pm0.11$ & $0.90\pm0.15$ \\
      \cmidrule{7-12}
      &  &  &  &  &  & $3_{12}-3_{03}$ & $1097$ & Ortho & $1500$ & $0.19\pm0.05$ & $0.16\pm0.04$ \\
      &  &  &  &  &  &  &  &  & $3100$ & $0.14\pm0.05$ & $0.13\pm0.05$ \\
      \cmidrule{7-12}
      &  &  &  &  &  & $1_{11}-0_{00}$ & $1113$ & Para & $1500$ & $1.53\pm0.07$ & $2.04\pm0.10$ \\
      &  &  &  &  &  &  &  &  & $3100$ & $0.67\pm0.08$ & $0.90\pm0.11$ \\   
    \midrule
    C/2009 P1 & 2011/10/16.79 & $-68.89$ & $1.807$ & $1.875$ & $1.36\pm0.22$ & $1_{10}-1_{01}$ & $557$ & Ortho & $69000$ & $255\pm10$ & $196\pm8$ \\
      \cmidrule{7-12}
      &  &  &  &  &  & $2_{11}-2_{02}$ & $752$ & Para & $69000$ & $22\pm6$ & $37\pm11$ \\
      \cmidrule{7-12}
      &  &  &  &  &  & $2_{02}-1_{11}$ & $988$ & Para & $45000$ & $110\pm6$ & $187\pm10$ \\
      &  &  &  &  &  &  &  &  & $69000$ & $97\pm3$ & $164\pm6$ \\
      &  &  &  &  &  &  &  &  & $90000$ & $87\pm14$ & $147\pm16$ \\
      \cmidrule{7-12}
      &  &  &  &  &  & $3_{12}-3_{03}$ & $1097$ & Ortho & $45000$ & $17\pm4$ & $13\pm3$ \\
      &  &  &  &  &  &  &  &  & $90000$ & $21\pm7$ & $16\pm6$ \\
      \cmidrule{7-12}
      &  &  &  &  &  & $1_{11}-0_{00}$ & $1113$ & Para & $45000$ & $115\pm5$ & $195\pm8$ \\
      &  &  &  &  &  &  &  &  & $90000$ & $69\pm12$ & $116\pm15$ \\     
    \midrule         
  \end{tabular}
  \label{tab:results}
\end{table*}

The $Q_{\rm{H_{2}O}}$ values were calculated by comparing the radiative transfer model and observed line intensities via a least-squared fitting method over a $Q_{\rm{H_{2}O}}$ range of 3 decades. $Q_{\rm{H_{2}O}}$ values were determined using the calculated OPR values shown in Table~\ref{tab:results}. Typical values of the OPR in comets are $2.5-3.0$ \citep{de_Val-Borro}, and so $Q_{\rm{H_{2}O}}$ values were calculated comets using an assumed OPR value of $3$ in order to draw comparisons. These are presented in Table~\ref{tab:results}.

The $Q_{\rm{H_{2}O}}$ values for Hartley 2 vary in the range $2-8\times10^{27}$\,s$^{-1}$ for the calculated OPR and $2-9\times10^{27}$\,s$^{-1}$ for an OPR $=3$. Aside from the value determined from the $3_{\rm{12}}-3_{\rm{03}}$ line, that is roughly an order of magnitude lower than for other transitions, these values are roughly equal to those calculated previously for observations taken at similar $r_{\rm{h}}$ as can be seen in Table~\ref{tab:comparison} \citep{Crovisier1999, DelloRusso2011, Combi2011b, Meech, Mumma2011, Kawakita, Knight, Gicquel}.

For Tempel 2 the $Q_{\rm{H_{2}O}}$ values presented here vary by an order of magnitude, $2-12\times10^{27}$\,s$^{-1}$ for the calculated OPR and $2-16\times10^{27}$\,s$^{-1}$ for OPR $=3$. By comparing these results with previous works Table~\ref{tab:comparison} shows that the upper values from these observations are a factor of $2-4$ lower than previously found \citep{Roettger, Fink, Szutowicz, Biver2012, Paganini2012Ic}.

As can be seen in Table~\ref{tab:comparison} the $Q_{\rm{H_{2}O}}$ values for 45P presented here, $0.6-1.4\times10^{27}$\,s$^{-1}$ and $0.9-2\times10^{27}$\,s$^{-1}$ for the calculated OPR and OPR $=3$ respectively, agree well with previous work observed at a similar time \citep{Fink, Lis2013ApJl}.

Values for $Q_{\rm{H_{2}O}}$ in the range $0.7-2.6\times10^{29}$\,s$^{-1}$ and $1.1-2.0\times10^{29}$\,s$^{-1}$ for the calculated OPR and OPR $=3$ respectively for C/2009 P1 were determined. These also agree well with the values presented in previous studies \citep{Bockelee-Morvan2012, Paganini2012ApJ, Villanueva, Combi2013, Bockelee-Morvan2014, DiSanti, Feaga}.

$Q_{\rm{H_{2}O}}$ values that are obtained from different transitions and offsets exhibit several features: (i) for Tempel 2, the rates calculated from all transitions are approximately similar, (ii) for Hartley 2, the rate determined from the $3_{\rm{12}}-3_{\rm{03}}$ line is significantly lower than observed from other transitions, (iii) for 45P and C/2009 P1, the rates calculated from the $3_{\rm{12}}-3_{\rm{03}}$ and $2_{\rm{11}}-2_{\rm{02}}$ lines are both noticeably lower than that determined from other transitions, and (iv) for all comets there is evidence of a trend for $Q_{\rm{H_{2}O}}$ to {\em decrease} with increasing offset.

For comets with an observed OPR lower than the canonical, statistical equilibrium value of $3$, the $Q_{\rm{H_{2}O}}$ values are higher for ortho-water lines and lower for para-water lines. For Hartley 2, whose calculated OPR is slightly higher than $3$, the reverse is true. This is as expected, because in order to reproduce the observed line intensities a higher ortho-$Q_{\rm{H_{2}O}}$ (and lower para-$Q_{\rm{H_{2}O}}$) would be needed if the OPR is lower than the statistical equilibrium value.

Significantly, for the majority of the observed comets, the use of these OPR values also results in a more consistent, narrower range of $Q_{\rm{H_{2}O}}$ values as determined from different transitions in each comet. This gives some support for the adoption of these OPR values. As can be seen in Table~\ref{tab:results}, even for low OPR values, the $Q_{\rm{H_{2}O}}$ values do not differ from those determined using an OPR $=3$ by more than a factor of two. 

\begin{table*}
  \caption{OPR and $Q_{\rm{H_{2}O}}$ values determined from observations taken at similar $r_{\rm{h}}$.} 
  \begin{threeparttable}
  \begin{tabular}{c c c c c c} 
    \toprule                   
    Comet & Apparition & $r_{\rm{h_{obs}}}$\,(au) & OPR & $Q_{\rm{H_{2}O}}$\,($10^{27}$\,s$^{-1}$) & References\\
    \bottomrule\toprule                  
    Hartley 2 & $2010$ & $1.07$ & $2.44\pm0.71$ & $1.89-7.46$ & $1$\\
     &  &  & 3 & $2.19-8.67$ & $1$ \\    
     \cmidrule{2-6}
     & $1991$ & $0.96$ & -- & $\approx63.00$ & $2$\\     
     &  & $1.05$ & -- & $32.36$ & $3$\\
     & $1997$ & $1.04$ & -- & $31.00\pm2.00$ & $4$\\   
     &  & $1.04$ & $2.76\pm0.08$ & $12.40\pm2.00$ & $5$\\
     & & $1.06$ & -- & $15.14$ & $3$\\          
     &  & $1.11$ & $2.63\pm0.18$ & $5.40\pm2.00$ & $5$\\      
     &  & $1.17$ & -- & $18.80$ & $6$\\     
     & $2010$ & $1.06$ & $3.4\pm0.6$ & $8.84-14.00$ & $7$\\
     &  & $1.06$ & $2.76\pm0.15$ & $8.44-13.60$ & $8$\\    
     &  & $1.06$ & -- & $11.48$ & $3$\\ 
     &  & $1.06$ & $2.85\pm0.20$ & $6.78\pm0.26$ & $9$\\                
     &  & $1.07$ & -- & $7.56\pm0.08$ & $10$\\     
     &  & $1.07$ & -- & $7.32\pm0.95$ & $11$\\     
     &  & $1.07$ & $2.88\pm0.17$ & $7.60-16.20$ & $8$\\
     &  & $1.07$ & -- & $\approx10.00$ & $12$\\
    \midrule
    Tempel 2 & $2010$ & $1.42$ & $1.59\pm0.23$ & $2.2-11.6$ & $1$ \\
    &  &  & 3 & $1.8-15.8$ & $1$ \\ 
    \cmidrule{2-6}  
    & $1988$ & $1.41$ & -- & $48.7$ & $6$\\
    & $1988$ & $1.42$  & -- & $\approx15.0-20.0$ & $13$\\  
    & $2010$ & $1.42$ & -- & $\approx20.0$ & $14$\\ 
    & $2010$ & $1.43$ & -- & $22.0\pm1.0$ & $15$\\
    & $2010$ & $1.44$ & $3.01\pm0.18$ & $19.0\pm1.2$ & $16$\\            
    \midrule
    45P & $2011$ & $1.00$ & $2.00\pm0.30$ & $0.60-1.36$ & $1$ \\
    &  &  & 3 & $0.90-2.04$ & $1$ \\     
    \cmidrule{2-6}
    & $1995$ & $1.14$ & -- & $1.92$ & $6$\\     
    & $2011$ & $1.03$ & -- & $0.91$ & $17$\\    
    \midrule
    C/2009 P1 & $2011$ & $1.81$ & $1.36\pm0.22$ & $69-255$ & $1$ \\
    &  &  & 3 & $114-196$ & $1$ \\     
    \cmidrule{2-6}
    & $2011$ & $1.73$ & -- & $108\pm30$ & $18$\\      
    &  & $1.76$ & -- & $69-81$ & $18$\\  
    &  & $1.80$ & -- & $270\pm3$ & $19$\\
    &  & $1.84$ & -- & $90-106$ & $20$\\ 
    &  & $1.88$ & -- & $155-262$ & $21$\\                        
    &  & $2.00$ & -- & $46\pm8$ & $22$\\                     
    &  & $2.00$ & -- & $84\pm7$ & $23$\\
    &  & $2.10$ & -- & $86\pm7$ & $24$\\     
    \midrule         
  \end{tabular}
  \begin{tablenotes}
   \item\textbf{References.} (1) This work; (2) \cite{Weaver}; (3) \cite{Knight}; (4) \cite{Colangeli}; (5) \cite{Crovisier1999}; (6) \cite{Fink}; (7) \cite{DelloRusso2011}; (8) \cite{Kawakita}; (9) \cite{Mumma2011}; (10) \cite{Combi2011b}; (11) \cite{Gicquel}; (12) \cite{Meech}; (13) \cite{Roettger}; (14) \cite{Szutowicz}; (15) \cite{Biver2012}; (16) \cite{Paganini2012Ic}; (17) \cite{Lis2013ApJl}; (18) \cite{Bockelee-Morvan2014}; (19) \cite{Combi2013}; (20) \cite{DiSanti}.; (21) \cite{Bockelee-Morvan2012}; (22) \cite{Feaga}; (23) \cite{Paganini2012ApJ}; (24) \cite{Villanueva}.
  \end{tablenotes}
  \end{threeparttable}
  \label{tab:comparison}
\end{table*}

\section{Discussion}
\label{sec:disc}

\subsection{OPR Variation}
\label{sec:opr_var}

One of the main results of this study is that the observed OPR of three of the comets, Tempel 2, 45P, \& C/2009 P1, presented in Table~\ref{tab:results}, are considerably lower than the statistical equilibrium value of $3$. However, as has been mentioned above, these values do agree with the OPR observed in other comets and interestingly they agree well with values found in the protoplanetary disc TW Hydrae that has a range of OPR from $0.73-1.52$ depending on the disc model \citep{Salinas}. Previously, an OPR of less than $3$ was thought to be due to a lower comet ice formation temperature, however recent laboratory studies have shown that solid-phase rapid nuclear-spin conversion equilibrates the OPR and gas-phase nuclear-spin conversion processes mentioned previously may be the cause of the OPR variation seen \citep{Hama2016}.

Previous observations of Tempel 2 were taken at lower nucleocentric distances than the observations presented here \citep{Paganini2012Ic}. The observations were centered on the nucleus, extending up to $1500$\,km. As previously mentioned, laboratory results seem to suggest that the Tempel 2 OPR previously determined is due to rapid nuclear-spin conversion normalising the OPR to the statistical equilibrium and the observed post-sublimation OPR value was 3. Whereas the OPR presented here was observed at a greater nucleocentric distance suggesting that nuclear-spin conversion via proton-transfer reactions of water with H$^{+}$ and H$_{3}$O$^{+}$ or water molecule collisions with water clusters, ice grains, or paramagnetic dust grains has occurred. 

It should also be noted that the Hartley 2 OPR values taken from the literature were determined from observations centered on-nucleus and extending up to several hundred km. All OPR values roughly agree with the statistical equilibrium OPR value of 3. For the SPIRE observations presented here, the on-nucleus OPR values for all comets in this study agree with the coma OPR value reported in Table~\ref{tab:results}. Assuming rapid nuclear-spin conversion occurs in the solid-phase on the nucleus, then due to the relatively large SPIRE beam sizes coma nuclear-spin conversion is being observed at nucleocentric distances less than $\sim1000-10000$\,km.

One of the first determinations of a cometary OPR was in comet C/1995 O1 (Hale--Bopp) in 1996 using ISO \citep{Crovisier1997}. An OPR of $2.45\pm0.10$ was reported using on-nucleus observations that extended up to roughly $20000$\,km suggesting that if the OPR of sublimated water should be equal to 3 then the ISO observation has also observed water that has undergone nuclear-spin conversion in the coma.

Interestingly, observations of the very inner coma of 73P-B/Schwassmann-Wachmann 3 and inner coma of C/2004 Q2 (Machholz) determined an OPR of roughly 3 and showed no variation over nucleocentric distances of $5-30$\,km and $\leq1000$\,km respectively \citep{Bonev2007, Bonev2008}. Furthermore, on-nucleus observations extending up to a nucleocentric distance of $350-1700$\,km taken of C/1999 H1 (Lee), C/1999 S4 (LINEAR), and C/2001 A2 (LINEAR) roughly agree with a statistical equilibrium OPR value. However, an OPR variation from $2.5-1.8$ was seen in C/2001 A2 (LINEAR) from observations at the same nucleocentric and similar heliocentric distances suggesting nuclear-spin conversion via the processes mentioned above was observed \citep{DelloRusso2005}.

It should also be noted that no significant difference in OPR for comets from the different comet families was seen. This is consistent with a recent study of ammonia in 26 comets \citep{Shinnaka}.

These observations seem to further support rapid solid-phase nuclear-spin conversion in cometary ice. However, these results highlight the need for further study of nuclear-spin conversion in cometary comae as it has previously been proposed that the processes mentioned above would occur in the very inner coma, but from the analysis above it would seem that OPR variation could occur at a range of nucleocentric distances.

\subsection{Water Production Rate Variation}

The variation of $Q_{\rm{H_{2}O}}$ with offset position is difficult to explain and there are a number of possible causes:
\begin{enumerate}
\item The temperature is not constant, but varies with position as line intensities are sensitive to the gas temperature profile in the inner coma.
\item There is an extra source of water in comae, perhaps from the sublimation of the dust ice mantles.
\item There is a spatial variation in the OPR.
\item The assumption of spherical symmetry in the radiative transfer model, both with respect to the outflow profile and the treatment of the photolysis reactions, may not be correct.
\item The excitation model for the water transitions may be over-simplified and/or missing key processes such as transitions from higher rotational levels.
\end{enumerate}
There is insufficient spatial resolution/data points to discriminate between these possibilities, but the strong discrepancy between the $Q_{\rm{H_{2}O}}$ values inferred from the $3_{\rm{12}}-3_{\rm{03}}$ (and $2_{\rm{11}}-2_{\rm{02}}$) lines and the other transitions suggests that an over-simplified model may be at least part of the cause of the observed variations. Furthermore, the pattern of the $Q_{\rm{H_{2}O}}$ variations varies from source to source and transition to transition, suggesting that temperature variation may not be the sole cause of the variation.

Note that the only Oort-cloud comet (C/2009 P1) in this study has a range of $Q_{\rm{H_{2}O}}$ values of at least an order of magnitude higher than the Jupiter-family comets. By comparing the determined $Q_{\rm{H_{2}O}}$ values in this study with other comets, C/2009 P1 has the second highest $Q_{\rm{H_{2}O}}$ at the observed $r_{\rm{h}}$. Comet C/1995 O1 (Hale--Bopp) was observed to have a $Q_{\rm{H_{2}O}}$ an order of magnitude higher \citep{Combi2000}. Interestingly, 45P has the lowest known $Q_{\rm{H_{2}O}}$ at the observed $r_{\rm{h}}$. Comet 67P/Churyumov--Gerasimenko has a similarly low $Q_{\rm{H_{2}O}}$ at a slightly greater $r_{\rm{h}}$ \citep{Bertaux}, although the observed values vary considerably with little change in $r_{\rm{h}}$. 

From the comparisons above and the fact that two comets in this study (Hartley 2 and Tempel 2) have lower $Q_{\rm{H_{2}O}}$ values than any Oort-cloud comet at the observed $r_{\rm{h}}$, one could draw the conclusion that Oort-cloud comets have greater $Q_{\rm{H_{2}O}}$ values due to a combination of a greater retention of accreted volatiles during formation and fewer perihelion passes. While in general this seems to be true, there are exceptions as 1P/Halley and 21P/Giacobini--Zinner have higher $Q_{\rm{H_{2}O}}$ values than C/2011 L4 (PanSTARRS) and C/2012 S1 (ISON) at similar observed $r_{\rm{h}}$ \citep{Combi1993, Combi2011a, Combi2014AJ, Combi2014apj}. Although this could be due to the nuclei of 1P/Halley and 21P/Giacobini--Zinner being larger than C/2012 S1 (ISON) and C/2011 L4 (PanSTARRS) \citep{Ferrin}. 

Furthermore, 19P/Borrelly has a $Q_{\rm{H_{2}O}}$ approximately a factor of 2 greater than C/1997 T1 (Utsunomiya) at a similar observed $r_{\rm{h}}$ \citep{Makinen, Combi2011a}. Although no definite  $r_{\rm{n}}$ for C/1997 T1 (Utsunomiya) has been calculated, an upper limit of $5.8$\,km has been determined \citep{Fernandez}, whereas 19P/Borrelly has a  $r_{\rm{n}}$ of $1.9$\,km \citep*{Lowry}. If the $r_{\rm{n}}$ of C/1997 T1 (Utsunomiya) is equal or greater to that of 19P/Borrelly then comet family might not influence $Q_{\rm{H_{2}O}}$, however if the nucleus radius of the Oort-cloud comet is much lower than $1.9$\,km then there would be more evidence that comet family affects $Q_{\rm{H_{2}O}}$.

By looking at Table~\ref{tab:comparison} another interesting point can be noted. The $Q_{\rm{H_{2}O}}$ values of Hartley 2, Tempel 2, and 45P for previous apparitions are higher at similar observed $r_{\rm{h}}$ than values presented in this study. Indeed, for Hartley 2 a difference in $Q_{\rm{H_{2}O}}$ after one orbit can be seen between the $1991$ and $1997$ apparitions at a $r_{\rm{h}}=1.04-1.06$\,au. For 45P the  $Q_{\rm{H_{2}O}}$ value during the $1995$ apparition is $\approx1.5-3$ times greater than the value reported in this study for a greater $r_{\rm{h}}$, therefore strengthening the idea that in general Oort-cloud comets have a greater $Q_{\rm{H_{2}O}}$ due to fewer perihelion passes. 

\section{Conclusions}
\label{sec:conc}

Using spectroscopic observations taken by \textit{Herschel}/SPIRE OPR values of three Jupiter-family comets and one Oort-cloud comet were determined. While the OPR for Hartley 2 is consistent with previous studies, the value calculated for Tempel 2 is lower than previously observed. The observed variation from the literature value could be due to gas-phase nuclear-spin conversion in the coma that occurred post OPR equilibration in the solid-phase on the nucleus and sublimation. The first OPR values for 45P and C/2009 P1 are presented. Although the OPR values for all comets aside from Hartley 2 are amongst the lowest observed, they agree with OPRs determined from previous observations of comets 1P/Halley and C/2001 A2, and the protoplanetary disc TW Hydrae. An important result of this study is that the observations are consistent with the findings of recent laboratory studies and that the OPR values presented provide good evidence of post-sublimation gas-phase nuclear-spin conversion.

From the OPR values the nuclear-spin-temperature of the four comets were determined and found that there is no substantial difference in the nuclear-spin-temperatures for comets from different families.

An established radiative transfer model and the calculated OPR values were used to determine $Q_{\rm{H_{2}O}}$ values, that vary with OPR as expected.

$Q_{\rm{H_{2}O}}$ values generally agree within an order of magnitude with previous observations, however notable exceptions are that the $Q_{\rm{H_{2}O}}$ values for all four comets determined from the $3_{\rm{12}}-3_{\rm{03}}$ ortho-water line (and for two comets, the $2_{\rm{11}}-2_{\rm{02}}$ para-water line) are lower. Consistently lower values across all four comets could suggest potential level population inaccuracies in the model, however further work is needed into this.

In general, the $Q_{\rm{H_{2}O}}$ values decrease with increasing offset from the comet. There are a number of possible explanations for this, but inaccuracies in the excitation model are potentially the most likely cause.

In this survey the only Oort-cloud comet, C/2009 P1, has a $Q_{\rm{H_{2}O}}$ value one-two orders of magnitude higher than the Jupiter-family comets. Placing the results reported here in context with the literature, C/2009 P1 has one of the highest $Q_{\rm{H_{2}O}}$ values at the observed $r_{\rm{h}}$ and that the three Jupiter-family comets have some of the lowest $Q_{\rm{H_{2}O}}$ values seen. One could conclude that $Q_{\rm{H_{2}O}}$ is related to comet family and therefore formation conditions. However, as mentioned previously there are potential exceptions that put this relationship into doubt. The $Q_{\rm{H_{2}O}}$ values for the three Jupiter-family comets in this study have been seen to decrease from previous apparitions suggesting Oort-cloud comets have a greater $Q_{\rm{H_{2}O}}$ due to fewer perihelion passes. \\

\section*{Acknowledgements}

The authors would like to thank Jay Farihi and Tetsuya Hama for comments and suggestions that greatly improved the paper. SPIRE has been developed by a consortium of institutes led by Cardiff University (UK) and including Univ. Lethbridge (Canada); NAOC (China); CEA, LAM (France); IFSI, Univ. Padua (Italy); IAC (Spain); Stockholm Observatory (Sweden); Imperial College London, RAL, UCL-MSSL, UKATC, Univ. Sussex (UK); and Caltech, JPL, NHSC, Univ. Colorado (USA). This development has been supported by national funding agencies: CSA (Canada); NAOC (China); CEA, CNES, CNRS (France); ASI (Italy); MCINN (Spain); SNSB (Sweden); STFC, UKSA (UK); and NASA (USA).
TGW wishes to acknowledge funding from a STFC studentship. 




\bibliographystyle{mnras}
\bibliography{Thomas_Wilson_Water_Production_Rates_and_Ortho_to_Para_Ratios} 

\begin{thebibliography}{}
\makeatletter
\relax
\def\mn@urlcharsother{\let\do\@makeother \do\$\do\&\do\#\do\^\do\_\do\%\do\~}
\def\mn@doi{\begingroup\mn@urlcharsother \@ifnextchar [ {\mn@doi@}
  {\mn@doi@[]}}
\def\mn@doi@[#1]#2{\def\@tempa{#1}\ifx\@tempa\@empty \href
  {http://dx.doi.org/#2} {doi:#2}\else \href {http://dx.doi.org/#2} {#1}\fi
  \endgroup}
\def\mn@eprint#1#2{\mn@eprint@#1:#2::\@nil}
\def\mn@eprint@arXiv#1{\href {http://arxiv.org/abs/#1} {{\tt arXiv:#1}}}
\def\mn@eprint@dblp#1{\href {http://dblp.uni-trier.de/rec/bibtex/#1.xml}
  {dblp:#1}}
\def\mn@eprint@#1:#2:#3:#4\@nil{\def\@tempa {#1}\def\@tempb {#2}\def\@tempc
  {#3}\ifx \@tempc \@empty \let \@tempc \@tempb \let \@tempb \@tempa \fi \ifx
  \@tempb \@empty \def\@tempb {arXiv}\fi \@ifundefined
  {mn@eprint@\@tempb}{\@tempb:\@tempc}{\expandafter \expandafter \csname
  mn@eprint@\@tempb\endcsname \expandafter{\@tempc}}}

\bibitem[\protect\citeauthoryear{Altwegg et~al.,}{Altwegg
  et~al.}{2015}]{Altwegg}
Altwegg K.,  et~al., 2015, \mn@doi [Science] {10.1126/science.1261952}, 347

\bibitem[\protect\citeauthoryear{{Balsiger}}{{Balsiger}}{1990}]{Balsiger}
{Balsiger} H.,  1990, {Measurements of ion species within the coma of comet
  Halley from Giotto.}.
pp 129--146

\bibitem[\protect\citeauthoryear{{Bensch} \& {Bergin}}{{Bensch} \&
  {Bergin}}{2004}]{Bensch}
{Bensch} F.,  {Bergin} E.~A.,  2004, \mn@doi [\apj] {10.1086/424439}, \href
  {http://adsabs.harvard.edu/abs/2004ApJ...615..531B} {615, 531}

\bibitem[\protect\citeauthoryear{{Bertaux}, {Combi}, {Qu{\'e}merais}  \&
  {Schmidt}}{{Bertaux} et~al.}{2014}]{Bertaux}
{Bertaux} J.-L.,  {Combi} M.~R.,  {Qu{\'e}merais} E.,   {Schmidt} W.,  2014,
  \mn@doi [\planss] {10.1016/j.pss.2013.11.006}, \href
  {http://adsabs.harvard.edu/abs/2014P%26SS...91...14B} {91, 14}

\bibitem[\protect\citeauthoryear{{Biver}}{{Biver}}{1997}]{Biver1997}
{Biver} N.,  1997, PhD thesis, Univ.~Paris 7-Diderot, (1997)

\bibitem[\protect\citeauthoryear{{Biver} et~al.,}{{Biver}
  et~al.}{2007}]{Biver2007}
{Biver} N.,  et~al., 2007, \mn@doi [\planss] {10.1016/j.pss.2006.11.010}, \href
  {http://adsabs.harvard.edu/abs/2007P%26SS...55.1058B} {55, 1058}

\bibitem[\protect\citeauthoryear{{Biver} et~al.,}{{Biver}
  et~al.}{2009}]{Biver2009}
{Biver} N.,  et~al., 2009, \mn@doi [\aap] {10.1051/0004-6361/200911790}, \href
  {http://adsabs.harvard.edu/abs/2009A%26A...501..359B} {501, 359}

\bibitem[\protect\citeauthoryear{{Biver} et~al.,}{{Biver}
  et~al.}{2012}]{Biver2012}
{Biver} N.,  et~al., 2012, \mn@doi [\aap] {10.1051/0004-6361/201118447}, \href
  {http://adsabs.harvard.edu/abs/2012A%26A...539A..68B} {539, A68}

\bibitem[\protect\citeauthoryear{{Bockel{\'e}e-Morvan}
  et~al.,}{{Bockel{\'e}e-Morvan} et~al.}{2012}]{Bockelee-Morvan2012}
{Bockel{\'e}e-Morvan} D.,  et~al., 2012, \mn@doi [\aap]
  {10.1051/0004-6361/201219744}, \href
  {http://adsabs.harvard.edu/abs/2012A%26A...544L..15B} {544, L15}

\bibitem[\protect\citeauthoryear{{Bockel{\'e}e-Morvan}
  et~al.,}{{Bockel{\'e}e-Morvan} et~al.}{2014}]{Bockelee-Morvan2014}
{Bockel{\'e}e-Morvan} D.,  et~al., 2014, \mn@doi [\aap]
  {10.1051/0004-6361/201322939}, \href
  {http://adsabs.harvard.edu/abs/2014A%26A...562A...5B} {562, A5}

\bibitem[\protect\citeauthoryear{{Bockel{\'e}e-Morvan}
  et~al.,}{{Bockel{\'e}e-Morvan} et~al.}{2015}]{Bockelee-Morvan2015}
{Bockel{\'e}e-Morvan} D.,  et~al., 2015, \mn@doi [\ssr]
  {10.1007/s11214-015-0156-9}, \href
  {http://adsabs.harvard.edu/abs/2015SSRv..197...47B} {197, 47}

\bibitem[\protect\citeauthoryear{{Bonev}, {Mumma}, {Villanueva}, {Disanti},
  {Ellis}, {Magee-Sauer}  \& {Dello Russo}}{{Bonev} et~al.}{2007}]{Bonev2007}
{Bonev} B.~P.,  {Mumma} M.~J.,  {Villanueva} G.~L.,  {Disanti} M.~A.,  {Ellis}
  R.~S.,  {Magee-Sauer} K.,   {Dello Russo} N.,  2007, \mn@doi [\apjl]
  {10.1086/518419}, \href {http://adsabs.harvard.edu/abs/2007ApJ...661L..97B}
  {661, L97}

\bibitem[\protect\citeauthoryear{{Bonev}, {Mumma}, {Kawakita}, {Kobayashi}  \&
  {Villanueva}}{{Bonev} et~al.}{2008}]{Bonev2008}
{Bonev} B.~P.,  {Mumma} M.~J.,  {Kawakita} H.,  {Kobayashi} H.,   {Villanueva}
  G.~L.,  2008, \mn@doi [\icarus] {10.1016/j.icarus.2008.02.023}, \href
  {http://adsabs.harvard.edu/abs/2008Icar..196..241B} {196, 241}

\bibitem[\protect\citeauthoryear{{Bonev}, {Villanueva}, {Paganini}, {DiSanti},
  {Gibb}, {Keane}, {Meech}  \& {Mumma}}{{Bonev} et~al.}{2013}]{Bonev2013}
{Bonev} B.~P.,  {Villanueva} G.~L.,  {Paganini} L.,  {DiSanti} M.~A.,  {Gibb}
  E.~L.,  {Keane} J.~V.,  {Meech} K.~J.,   {Mumma} M.~J.,  2013, \mn@doi
  [\icarus] {10.1016/j.icarus.2012.07.034}, \href
  {http://adsabs.harvard.edu/abs/2013Icar..222..740B} {222, 740}

\bibitem[\protect\citeauthoryear{Buntkowsky et~al.,}{Buntkowsky
  et~al.}{2008}]{Buntkowsky}
Buntkowsky G.,  et~al., 2008, Zeitschrift f{\"u}r Physikalische Chemie
  International journal of research in physical chemistry and chemical physics,
  222, 1049

\bibitem[\protect\citeauthoryear{{Choi}, {van der Tak}, {Bergin}  \&
  {Plume}}{{Choi} et~al.}{2014}]{Choi}
{Choi} Y.,  {van der Tak} F.~F.~S.,  {Bergin} E.~A.,   {Plume} R.,  2014,
  \mn@doi [\aap] {10.1051/0004-6361/201424007}, \href
  {http://adsabs.harvard.edu/abs/2014A%26A...572L..10C} {572, L10}

\bibitem[\protect\citeauthoryear{{Colangeli} et~al.,}{{Colangeli}
  et~al.}{1999}]{Colangeli}
{Colangeli} L.,  et~al., 1999, \aap, \href
  {http://adsabs.harvard.edu/abs/1999A%26A...343L..87C} {343, L87}

\bibitem[\protect\citeauthoryear{{Combi} \& {Feldman}}{{Combi} \&
  {Feldman}}{1993}]{Combi1993}
{Combi} M.~R.,  {Feldman} P.~D.,  1993, \mn@doi [\icarus]
  {10.1006/icar.1993.1149}, \href
  {http://adsabs.harvard.edu/abs/1993Icar..105..557C} {105, 557}

\bibitem[\protect\citeauthoryear{{Combi}, {Cochran}, {Cochran}, {Lambert}  \&
  {Johns-Krull}}{{Combi} et~al.}{1999}]{Combi1999}
{Combi} M.~R.,  {Cochran} A.~L.,  {Cochran} W.~D.,  {Lambert} D.~L.,
  {Johns-Krull} C.~M.,  1999, \mn@doi [\apj] {10.1086/306798}, \href
  {http://adsabs.harvard.edu/abs/1999ApJ...512..961C} {512, 961}

\bibitem[\protect\citeauthoryear{{Combi}, {Reinard}, {Bertaux}, {Quemerais}  \&
  {M{\"a}kinen}}{{Combi} et~al.}{2000}]{Combi2000}
{Combi} M.~R.,  {Reinard} A.~A.,  {Bertaux} J.-L.,  {Quemerais} E.,
  {M{\"a}kinen} T.,  2000, \mn@doi [\icarus] {10.1006/icar.1999.6335}, \href
  {http://adsabs.harvard.edu/abs/2000Icar..144..191C} {144, 191}

\bibitem[\protect\citeauthoryear{{Combi}, {Lee}, {Patel}, {M{\"a}kinen},
  {Bertaux}  \& {Qu{\'e}merais}}{{Combi} et~al.}{2011a}]{Combi2011a}
{Combi} M.~R.,  {Lee} Y.,  {Patel} T.~S.,  {M{\"a}kinen} J.~T.~T.,  {Bertaux}
  J.-L.,   {Qu{\'e}merais} E.,  2011a, \mn@doi [\aj]
  {10.1088/0004-6256/141/4/128}, \href
  {http://adsabs.harvard.edu/abs/2011AJ....141..128C} {141, 128}

\bibitem[\protect\citeauthoryear{{Combi}, {Bertaux}, {Qu{\'e}merais}, {Ferron}
  \& {M{\"a}kinen}}{{Combi} et~al.}{2011b}]{Combi2011b}
{Combi} M.~R.,  {Bertaux} J.-L.,  {Qu{\'e}merais} E.,  {Ferron} S.,
  {M{\"a}kinen} J.~T.~T.,  2011b, \mn@doi [\apjl] {10.1088/2041-8205/734/1/L6},
  \href {http://adsabs.harvard.edu/abs/2011ApJ...734L...6C} {734, L6}

\bibitem[\protect\citeauthoryear{{Combi}, {M{\"a}kinen}, {Bertaux},
  {Qu{\'e}merais}, {Ferron}  \& {Fougere}}{{Combi} et~al.}{2013}]{Combi2013}
{Combi} M.~R.,  {M{\"a}kinen} J.~T.~T.,  {Bertaux} J.-L.,  {Qu{\'e}merais} E.,
  {Ferron} S.,   {Fougere} N.,  2013, \mn@doi [\icarus]
  {10.1016/j.icarus.2013.04.030}, \href
  {http://adsabs.harvard.edu/abs/2013Icar..225..740C} {225, 740}

\bibitem[\protect\citeauthoryear{{Combi}, {Bertaux}, {Qu{\'e}merais}, {Ferron},
  {M{\"a}kinen}  \& {Aptekar}}{{Combi} et~al.}{2014a}]{Combi2014AJ}
{Combi} M.~R.,  {Bertaux} J.-L.,  {Qu{\'e}merais} E.,  {Ferron} S.,
  {M{\"a}kinen} J.~T.~T.,   {Aptekar} G.,  2014a, \mn@doi [\aj]
  {10.1088/0004-6256/147/6/126}, \href
  {http://adsabs.harvard.edu/abs/2014AJ....147..126C} {147, 126}

\bibitem[\protect\citeauthoryear{{Combi}, {Fougere}, {M{\"a}kinen}, {Bertaux},
  {Qu{\'e}merais}  \& {Ferron}}{{Combi} et~al.}{2014b}]{Combi2014apj}
{Combi} M.~R.,  {Fougere} N.,  {M{\"a}kinen} J.~T.~T.,  {Bertaux} J.-L.,
  {Qu{\'e}merais} E.,   {Ferron} S.,  2014b, \mn@doi [\apjl]
  {10.1088/2041-8205/788/1/L7}, \href
  {http://adsabs.harvard.edu/abs/2014ApJ...788L...7C} {788, L7}

\bibitem[\protect\citeauthoryear{{Crovisier}}{{Crovisier}}{1984}]{Crovisier1984}
{Crovisier} J.,  1984, \aap, \href
  {http://adsabs.harvard.edu/abs/1984A%26A...130..361C} {130, 361}

\bibitem[\protect\citeauthoryear{{Crovisier}, {Leech}, {Bockelee-Morvan},
  {Brooke}, {Hanner}, {Altieri}, {Keller}  \& {Lellouch}}{{Crovisier}
  et~al.}{1997}]{Crovisier1997}
{Crovisier} J.,  {Leech} K.,  {Bockelee-Morvan} D.,  {Brooke} T.~Y.,  {Hanner}
  M.~S.,  {Altieri} B.,  {Keller} H.~U.,   {Lellouch} E.,  1997, \mn@doi
  [Science] {10.1126/science.275.5308.1904}, \href
  {http://adsabs.harvard.edu/abs/1997Sci...275.1904C} {275, 1904}

\bibitem[\protect\citeauthoryear{{Crovisier} et~al.,}{{Crovisier}
  et~al.}{1999}]{Crovisier1999}
{Crovisier} J.,  et~al., 1999, in {Cox} P.,  {Kessler} M.,  eds,  ESA Special
  Publication Vol. 427, The Universe as Seen by ISO. p.~161

\bibitem[\protect\citeauthoryear{{De Val-Borro} et~al.,}{{De Val-Borro}
  et~al.}{2010}]{de_Val-Borro}
{De Val-Borro} M.,  et~al., 2010, \mn@doi [\aap] {10.1051/0004-6361/201015161},
  \href {http://adsabs.harvard.edu/abs/2010A%26A...521L..50D} {521, L50}

\bibitem[\protect\citeauthoryear{{Dello Russo}, {Bonev}, {DiSanti}, {Mumma},
  {Gibb}, {Magee-Sauer}, {Barber}  \& {Tennyson}}{{Dello Russo}
  et~al.}{2005}]{DelloRusso2005}
{Dello Russo} N.,  {Bonev} B.~P.,  {DiSanti} M.~A.,  {Mumma} M.~J.,  {Gibb}
  E.~L.,  {Magee-Sauer} K.,  {Barber} R.~J.,   {Tennyson} J.,  2005, \mn@doi
  [\apj] {10.1086/427473}, \href
  {http://adsabs.harvard.edu/abs/2005ApJ...621..537D} {621, 537}

\bibitem[\protect\citeauthoryear{{Dello Russo} et~al.,}{{Dello Russo}
  et~al.}{2011}]{DelloRusso2011}
{Dello Russo} N.,  et~al., 2011, \mn@doi [\apjl] {10.1088/2041-8205/734/1/L8},
  \href {http://adsabs.harvard.edu/abs/2011ApJ...734L...8D} {734, L8}

\bibitem[\protect\citeauthoryear{{DiSanti}, {Villanueva}, {Paganini}, {Bonev},
  {Keane}, {Meech}  \& {Mumma}}{{DiSanti} et~al.}{2014}]{DiSanti}
{DiSanti} M.~A.,  {Villanueva} G.~L.,  {Paganini} L.,  {Bonev} B.~P.,  {Keane}
  J.~V.,  {Meech} K.~J.,   {Mumma} M.~J.,  2014, \mn@doi [\icarus]
  {10.1016/j.icarus.2013.09.001}, \href
  {http://adsabs.harvard.edu/abs/2014Icar..228..167D} {228, 167}

\bibitem[\protect\citeauthoryear{{Feaga} et~al.,}{{Feaga} et~al.}{2014}]{Feaga}
{Feaga} L.~M.,  et~al., 2014, \mn@doi [\aj] {10.1088/0004-6256/147/1/24}, \href
  {http://adsabs.harvard.edu/abs/2014AJ....147...24F} {147, 24}

\bibitem[\protect\citeauthoryear{{Fernandez}}{{Fernandez}}{1999}]{Fernandez}
{Fernandez} Y.~R.,  1999, PhD thesis, University of Maryland College Park

\bibitem[\protect\citeauthoryear{{Ferr{\'{\i}}n}}{{Ferr{\'{\i}}n}}{2014}]{Ferrin}
{Ferr{\'{\i}}n} I.,  2014, \mn@doi [\mnras] {10.1093/mnras/stu820}, \href
  {http://adsabs.harvard.edu/abs/2014MNRAS.442.1731F} {442, 1731}

\bibitem[\protect\citeauthoryear{{Festou}}{{Festou}}{1990}]{Festou}
{Festou} M.~C.,  1990, {Variations of the gaseous output of the nucleus of
  comet Halley.}.
pp 245--257

\bibitem[\protect\citeauthoryear{{Fink}}{{Fink}}{2009}]{Fink}
{Fink} U.,  2009, \mn@doi [\icarus] {10.1016/j.icarus.2008.12.044}, \href
  {http://adsabs.harvard.edu/abs/2009Icar..201..311F} {201, 311}

\bibitem[\protect\citeauthoryear{{Gicquel} et~al.,}{{Gicquel}
  et~al.}{2014}]{Gicquel}
{Gicquel} A.,  et~al., 2014, \mn@doi [\apj] {10.1088/0004-637X/794/1/1}, \href
  {http://adsabs.harvard.edu/abs/2014ApJ...794....1G} {794, 1}

\bibitem[\protect\citeauthoryear{{Griffin} et~al.,}{{Griffin}
  et~al.}{2010}]{Griffin}
{Griffin} M.~J.,  et~al., 2010, \mn@doi [\aap] {10.1051/0004-6361/201014519},
  \href {http://adsabs.harvard.edu/abs/2010A%26A...518L...3G} {518, L3}

\bibitem[\protect\citeauthoryear{{Gulkis}}{{Gulkis}}{2014}]{Gulkis}
{Gulkis} S.,  2014, Central Bureau Electronic Telegrams, \href
  {http://adsabs.harvard.edu/abs/2014CBET.3912....1G} {3912}

\bibitem[\protect\citeauthoryear{{Hama} \& {Watanabe}}{{Hama} \&
  {Watanabe}}{2013}]{Hama2013}
{Hama} T.,  {Watanabe} N.,  2013, \mn@doi [Chemical Reviews]
  {10.1021/cr4000978}, \href
  {http://adsabs.harvard.edu/abs/2013ChRv..113.8783H} {113, 8783}

\bibitem[\protect\citeauthoryear{{Hama}, {Kouchi}  \& {Watanabe}}{{Hama}
  et~al.}{2016}]{Hama2016}
{Hama} T.,  {Kouchi} A.,   {Watanabe} N.,  2016, \mn@doi [Science]
  {10.1126/science.aad4026}, \href
  {http://adsabs.harvard.edu/abs/2016Sci...351...65H} {351, 65}

\bibitem[\protect\citeauthoryear{{Hartogh} et~al.,}{{Hartogh}
  et~al.}{2009}]{Hartogh2009}
{Hartogh} P.,  et~al., 2009, \mn@doi [\planss] {10.1016/j.pss.2009.07.009},
  \href {http://adsabs.harvard.edu/abs/2009P%26SS...57.1596H} {57, 1596}

\bibitem[\protect\citeauthoryear{{Hartogh} et~al.,}{{Hartogh}
  et~al.}{2010}]{Hartogh2010}
{Hartogh} P.,  et~al., 2010, \mn@doi [\aap] {10.1051/0004-6361/201014665},
  \href {http://adsabs.harvard.edu/abs/2010A%26A...518L.150H} {518, L150}

\bibitem[\protect\citeauthoryear{{Haser}}{{Haser}}{1957}]{Haser}
{Haser} L.,  1957, Bulletin de la Societe Royale des Sciences de Liege, \href
  {http://adsabs.harvard.edu/abs/1957BSRSL..43..740H} {43, 740}

\bibitem[\protect\citeauthoryear{{\textit{Herschel} Science
  Centre.}}{{\textit{Herschel} Science Centre.}}{2014}]{Herschel}
{\textit{Herschel} Science Centre.} 2014, {SPIRE Handbook 2016,
  \textit{Herschel} Explanatory Supplement vol. 2.5}.
HERSCHEL-DOC-0798

\bibitem[\protect\citeauthoryear{{Hogerheijde} \& {van der Tak}}{{Hogerheijde}
  \& {van der Tak}}{2000}]{Hogerheijde}
{Hogerheijde} M.~R.,  {van der Tak} F.~F.~S.,  2000, \aap, \href
  {http://adsabs.harvard.edu/abs/2000A%26A...362..697H} {362, 697}

\bibitem[\protect\citeauthoryear{{Irvine}, {Schloerb}, {Crovisier}, {Fegley}
  \& {Mumma}}{{Irvine} et~al.}{2000}]{Irvine}
{Irvine} W.~M.,  {Schloerb} F.~P.,  {Crovisier} J.,  {Fegley} Jr. B.,   {Mumma}
  M.~J.,  2000, Protostars and Planets IV, \href
  {http://adsabs.harvard.edu/abs/2000prpl.conf.1159I} {p.~1159}

\bibitem[\protect\citeauthoryear{{Kawakita} et~al.,}{{Kawakita}
  et~al.}{2013}]{Kawakita}
{Kawakita} H.,  et~al., 2013, \mn@doi [\icarus] {10.1016/j.icarus.2012.08.006},
  \href {http://adsabs.harvard.edu/abs/2013Icar..222..723K} {222, 723}

\bibitem[\protect\citeauthoryear{{Knight} \& {Schleicher}}{{Knight} \&
  {Schleicher}}{2013}]{Knight}
{Knight} M.~M.,  {Schleicher} D.~G.,  2013, \mn@doi [\icarus]
  {10.1016/j.icarus.2012.06.004}, \href
  {http://adsabs.harvard.edu/abs/2013Icar..222..691K} {222, 691}

\bibitem[\protect\citeauthoryear{{Lecacheux} et~al.,}{{Lecacheux}
  et~al.}{2003}]{Lecacheux}
{Lecacheux} A.,  et~al., 2003, \mn@doi [\aap] {10.1051/0004-6361:20030338},
  \href {http://adsabs.harvard.edu/abs/2003A%26A...402L..55L} {402, L55}

\bibitem[\protect\citeauthoryear{{Lis}, {Bergin}, {Schilke}  \& {van
  Dishoeck}}{{Lis} et~al.}{2013a}]{Lis2013PhysChem}
{Lis} D.~C.,  {Bergin} E.~A.,  {Schilke} P.,   {van Dishoeck} E.~F.,  2013a,
  \mn@doi [Journal of Physical Chemistry A] {10.1021/jp312333n}, \href
  {http://adsabs.harvard.edu/abs/2013JPCA..117.9661L} {117, 9661}

\bibitem[\protect\citeauthoryear{{Lis} et~al.,}{{Lis}
  et~al.}{2013b}]{Lis2013ApJl}
{Lis} D.~C.,  et~al., 2013b, \mn@doi [\apjl] {10.1088/2041-8205/774/1/L3},
  \href {http://adsabs.harvard.edu/abs/2013ApJ...774L...3L} {774, L3}

\bibitem[\protect\citeauthoryear{{Lowry}, {Fitzsimmons}  \&
  {Collander-Brown}}{{Lowry} et~al.}{2003}]{Lowry}
{Lowry} S.~C.,  {Fitzsimmons} A.,   {Collander-Brown} S.,  2003, \mn@doi [\aap]
  {10.1051/0004-6361:20021486}, \href
  {http://adsabs.harvard.edu/abs/2003A%26A...397..329L} {397, 329}

\bibitem[\protect\citeauthoryear{{M{\"a}kinen}, {Bertaux}, {Pulkkinen},
  {Schmidt}, {Kyr{\"o}l{\"a}}, {Summanen}, {Qu{\'e}merais}  \&
  {Lallement}}{{M{\"a}kinen} et~al.}{2001}]{Makinen}
{M{\"a}kinen} J.~T.~T.,  {Bertaux} J.-L.,  {Pulkkinen} T.~I.,  {Schmidt} W.,
  {Kyr{\"o}l{\"a}} E.,  {Summanen} T.,  {Qu{\'e}merais} E.,   {Lallement} R.,
  2001, \mn@doi [\aap] {10.1051/0004-6361:20000545}, \href
  {http://adsabs.harvard.edu/abs/2001A%26A...368..292M} {368, 292}

\bibitem[\protect\citeauthoryear{Manca~Tanner, Quack  \&
  Schmidiger}{Manca~Tanner et~al.}{2013}]{MancaTanner}
Manca~Tanner C.,  Quack M.,   Schmidiger D.,  2013, The Journal of Physical
  Chemistry A, 117, 10105

\bibitem[\protect\citeauthoryear{{Meech} et~al.,}{{Meech} et~al.}{2011}]{Meech}
{Meech} K.~J.,  et~al., 2011, \mn@doi [\apjl] {10.1088/2041-8205/734/1/L1},
  \href {http://adsabs.harvard.edu/abs/2011ApJ...734L...1M} {734, L1}

\bibitem[\protect\citeauthoryear{{Mumma}, {Weaver}  \& {Larson}}{{Mumma}
  et~al.}{1987}]{Mumma1987}
{Mumma} M.~J.,  {Weaver} H.~A.,   {Larson} H.~P.,  1987, \aap, \href
  {http://adsabs.harvard.edu/abs/1987A%26A...187..419M} {187, 419}

\bibitem[\protect\citeauthoryear{{Mumma}, {Blass}, {Weaver}  \&
  {Larson}}{{Mumma} et~al.}{1988}]{Mumma1988}
{Mumma} M.~J.,  {Blass} W.~E.,  {Weaver} H.~A.,   {Larson} H.~P.,  1988, in
  Bulletin of the American Astronomical Society. p.~826

\bibitem[\protect\citeauthoryear{{Mumma} et~al.,}{{Mumma}
  et~al.}{2011}]{Mumma2011}
{Mumma} M.~J.,  et~al., 2011, \mn@doi [\apjl] {10.1088/2041-8205/734/1/L7},
  \href {http://adsabs.harvard.edu/abs/2011ApJ...734L...7M} {734, L7}

\bibitem[\protect\citeauthoryear{{Neufeld} et~al.,}{{Neufeld}
  et~al.}{2000}]{Neufeld}
{Neufeld} D.~A.,  et~al., 2000, \mn@doi [\apjl] {10.1086/312852}, \href
  {http://adsabs.harvard.edu/abs/2000ApJ...539L.107N} {539, L107}

\bibitem[\protect\citeauthoryear{{Paganini}, {Mumma}, {Bonev}, {Villanueva},
  {DiSanti}, {Keane}  \& {Meech}}{{Paganini} et~al.}{2012a}]{Paganini2012Ic}
{Paganini} L.,  {Mumma} M.~J.,  {Bonev} B.~P.,  {Villanueva} G.~L.,  {DiSanti}
  M.~A.,  {Keane} J.~V.,   {Meech} K.~J.,  2012a, \mn@doi [\icarus]
  {10.1016/j.icarus.2012.01.004}, \href
  {http://adsabs.harvard.edu/abs/2012Icar..218..644P} {218, 644}

\bibitem[\protect\citeauthoryear{{Paganini}, {Mumma}, {Villanueva}, {DiSanti},
  {Bonev}, {Lippi}  \& {Boehnhardt}}{{Paganini}
  et~al.}{2012b}]{Paganini2012ApJ}
{Paganini} L.,  {Mumma} M.~J.,  {Villanueva} G.~L.,  {DiSanti} M.~A.,  {Bonev}
  B.~P.,  {Lippi} M.,   {Boehnhardt} H.,  2012b, \mn@doi [\apjl]
  {10.1088/2041-8205/748/1/L13}, \href
  {http://adsabs.harvard.edu/abs/2012ApJ...748L..13P} {748, L13}

\bibitem[\protect\citeauthoryear{{Pilbratt} et~al.,}{{Pilbratt}
  et~al.}{2010}]{Pilbratt}
{Pilbratt} G.~L.,  et~al., 2010, \mn@doi [\aap] {10.1051/0004-6361/201014759},
  \href {http://adsabs.harvard.edu/abs/2010A%26A...518L...1P} {518, L1}

\bibitem[\protect\citeauthoryear{{Roettger}, {Feldman}, {A'Hearn}  \&
  {Festou}}{{Roettger} et~al.}{1990}]{Roettger}
{Roettger} E.~E.,  {Feldman} P.~D.,  {A'Hearn} M.~F.,   {Festou} M.~C.,  1990,
  \mn@doi [\icarus] {10.1016/0019-1035(90)90202-K}, \href
  {http://adsabs.harvard.edu/abs/1990Icar...86..100R} {86, 100}

\bibitem[\protect\citeauthoryear{{Salinas} et~al.,}{{Salinas}
  et~al.}{2016}]{Salinas}
{Salinas} V.~N.,  et~al., 2016, \mn@doi [\aap] {10.1051/0004-6361/201628172},
  \href {http://adsabs.harvard.edu/abs/2016A%26A...591A.122S} {591, A122}

\bibitem[\protect\citeauthoryear{Shinnaka, Kawakita, Jehin, Decock,
  Hutsemékers  \& Manfroid}{Shinnaka et~al.}{2016}]{Shinnaka}
Shinnaka Y.,  Kawakita H.,  Jehin E.,  Decock A.,  Hutsemékers D.,   Manfroid
  J.,  2016, \mn@doi [Monthly Notices of the Royal Astronomical Society]
  {10.1093/mnras/stw2298}, 462, S124

\bibitem[\protect\citeauthoryear{{Swinyard} et~al.,}{{Swinyard}
  et~al.}{2014}]{Swinyard}
{Swinyard} B.~M.,  et~al., 2014, \mn@doi [\mnras] {10.1093/mnras/stu409}, \href
  {http://adsabs.harvard.edu/abs/2014MNRAS.440.3658S} {440, 3658}

\bibitem[\protect\citeauthoryear{{Szutowicz} et~al.,}{{Szutowicz}
  et~al.}{2011}]{Szutowicz}
{Szutowicz} S.,  et~al., 2011, in EPSC-DPS Joint Meeting 2011. p.~1213

\bibitem[\protect\citeauthoryear{{Villanueva}, {Mumma}, {DiSanti}, {Bonev},
  {Paganini}  \& {Blake}}{{Villanueva} et~al.}{2012}]{Villanueva}
{Villanueva} G.~L.,  {Mumma} M.~J.,  {DiSanti} M.~A.,  {Bonev} B.~P.,
  {Paganini} L.,   {Blake} G.~A.,  2012, \mn@doi [\icarus]
  {10.1016/j.icarus.2012.03.027}, \href
  {http://adsabs.harvard.edu/abs/2012Icar..220..291V} {220, 291}

\bibitem[\protect\citeauthoryear{{Weaver}, {Feldman}, {McPhate}, {A'Hearn},
  {Arpigny}  \& {Smith}}{{Weaver} et~al.}{1994}]{Weaver}
{Weaver} H.~A.,  {Feldman} P.~D.,  {McPhate} J.~B.,  {A'Hearn} M.~F.,
  {Arpigny} C.,   {Smith} T.~E.,  1994, \mn@doi [\apj] {10.1086/173732}, \href
  {http://adsabs.harvard.edu/abs/1994ApJ...422..374W} {422, 374}

\bibitem[\protect\citeauthoryear{{Willacy} et~al.,}{{Willacy}
  et~al.}{2015}]{Willacy}
{Willacy} K.,  et~al., 2015, \mn@doi [\ssr] {10.1007/s11214-015-0167-6}, \href
  {http://adsabs.harvard.edu/abs/2015SSRv..197..151W} {197, 151}

\bibitem[\protect\citeauthoryear{{Zakharov}, {Bockel{\'e}e-Morvan}, {Biver},
  {Crovisier}  \& {Lecacheux}}{{Zakharov} et~al.}{2007}]{Zakharov}
{Zakharov} V.,  {Bockel{\'e}e-Morvan} D.,  {Biver} N.,  {Crovisier} J.,
  {Lecacheux} A.,  2007, \mn@doi [\aap] {10.1051/0004-6361:20066715}, \href
  {http://adsabs.harvard.edu/abs/2007A%26A...473..303Z} {473, 303}

\makeatother
\end{thebibliography}




\appendix


\bsp	
\label{lastpage}
\end{document}